\documentclass[lettersize,journal]{IEEEtran}
\usepackage{amsmath,amsfonts}
\usepackage{algorithmic}
\usepackage{algorithm}
\usepackage{array}
\usepackage[caption=false,font=normalsize,labelfont=sf,textfont=sf]{subfig}
\usepackage{textcomp}
\usepackage{stfloats}
\usepackage{url}
\usepackage{verbatim}
\usepackage{graphicx}
\usepackage{cite}
\usepackage{multirow}
\graphicspath{{figures/}}

\begin{document}

\title{ESupNNet: An Error Supervising Neural Network architecture for error detection against soft errors in parameters}

\author{
	\IEEEauthorblockN{Jorge Cano-Páez\IEEEauthorrefmark{1}\thanks{This work has been supported by the Community of Madrid, under grant PIPF-2024/TEC-34925.}, Luis Entrena\IEEEauthorrefmark{2}, and Almudena Lindoso\IEEEauthorrefmark{3}\\}
	\IEEEauthorblockA{
		Department of Electronic Technology \\
		Universidad Carlos III de Madrid \\
		Leganés, Spain \\
		Email: \IEEEauthorrefmark{1}jocanop@ing.uc3m.es, \IEEEauthorrefmark{2}entrena@ing.uc3m.es, \IEEEauthorrefmark{3}alindoso@ing.uc3m.es
	}
}



\maketitle

\begin{abstract}
This work presents a novel approach to detect misclassification errors in CNNs caused by soft errors in their parameters. We propose an architecture that uses inter-class relations induced by the CNN that needs protection. The architecture has minimal resources overhead and does not require modifying the CNN, which makes it a competent solution that can be used with other error protection techniques. We have validated the architecture with five different combinations of modern dataset-model pairs: ImageNet-1K for ResNet-50 and EfficientNetV2-Small; CIFAR-10 for MobileNetV3, ShuffleNetV2-Small with 2.0x output channels and MNASNet with depth multiplier of 1.3. The validation process was done rigorously with statistical significance, from the creation of the datasets used by the architecture to the acquisition of experimental results. Results show great performance with over 90\% accuracy in detecting single errors and great error detection over multiple Bit Error Rates, which can be potentially increased with hyperparameter tuning.
\end{abstract}

\begin{IEEEkeywords}
Reliability, Fault tolerant, Neural network.
\end{IEEEkeywords}

\begin{table}[htbp]
	\caption{Nomenclature of Symbols}
	\label{tab:symbols}
	\centering
	\begin{tabular}{|l|l|}
		
		\hline
		\textbf{Symbol} & \textbf{Description}  \\
		\hline
		
		$D$ & Dataset used to train the supervising ANN, it is balanced \\
		$D_p$ & Parent dataset obtained by sampling using fault injection \\
		$E$ & Expected value \\
		$e$ & Error margin \\
		$F_{\theta_c}$ & Fully connected part of a CNN \\
		$f$ & Layer of the fully connected part of a CNN \\
		$M$ & Target CNN \\
		$n_0$ & Number of injections performed \\
		$p$ & Proportion of an attribute that is present in the population \\
		$R$ & Relation between classes \\
		$R^*$ & Approximation to relation between classes $R$ \\
		$S$ & Supervising ANN \\
		$X$ & Set of input data  \\
		$X_c$ & Set of $z$ for data correctly classified \\
		$X_{e}$ & Set of $z'$ for data misclassified \\
		$X_{s}$ & Set of $z'$ for data correctly classified \\
		$X_{*b}$ & Set of $z'$ for data with altered biases \\
		$X_{*w}$ & Set of $z'$ for data with altered weights \\
		$x$ & Input sample, element from $X$ \\
		$Y$ & Set of possible classes \\
		$y$ & Ground truth class, element from $Y$ \\
		$\hat{y}$ & Predicted class \\
		$\hat{y}'$ & Predicted class when parameters are altered by an error \\
		$Z$ & Z-score of a confidence interval \\
		$z$ & Vector of output logits of a CNN \\
		$z'$ & Vector of output logits of a CNN with altered parameters \\		
		
		$\delta$ & Kronecker delta \\
		$\theta_f$ & Set of parameters for the feature extractor of a CNN \\
		$\theta_c$ & Set of parameters for the fully connected part of a CNN \\
		$\theta'$ & Set of parameters altered by an error \\
		$\mu_i$ & Centroid of class $i$ \\
		$\sigma$ & Sigmoid function \\
		$\Phi_{\theta_f}$ & Feature extractor of a CNN \\
		
		$s(\cdot,\cdot)$ & Similarity measure \\
		
		\hline
		
	\end{tabular}
\end{table}

\section{Introduction}
\IEEEPARstart{T}{here} is an increasing interest in applying Artificial Neural Networks (ANNs) for a wide variety of applications such as medical image analysis \cite{med_0,med_1}, finance \cite{fina_0} and computer vision \cite{compv_0}. This interest is mostly due to their capability of modeling complex systems and solving classification and regression tasks in a cost-effective manner. In image processing applications, Convolutional Neural Networks (CNNs) are the backbone of most ANN implementations. CNNs are a specialised architecture that use kernels or filters for automatic feature extraction of input images. Some of the applications that use CNNs belong to the high reliability field, like autonomous driving \cite{drive_0,drive_1,drive_2} and unmanned navigation \cite{uav_0,uav_1}. For space applications, performing data processing on-board of satellites significantly reduces the amount of data that needs to be sent to ground stations. Moreover, developing computer vision based processes further improves the autonomy of future deep space exploration missions. CNNs are a suitable option for Earth observation \cite{earthobv_0,earthobv_1}, satellite imagery \cite{spaceimg_0,spaceimg_1} and space situational awareness \cite{spaceaware_0} using optical sensors.

CNNs have sensitive elements like feature maps, weights and biases that, if affected by an error, could make the system produce an erroneous inference. As the number of parameters increases so does the susceptibility of suffering errors, and modern CNNs have millions of parameters. An error affecting a CNN in an autonomous driving system can threat the life of people. In space applications like Earth observation or satellite imagery, an error can make the system to incorrectly infer objects, which could incur in a potentially loss of relevant data. Furthermore, in deep space exploration, an error can be the difference between the loss of a probe and the success of a mission, as autonomous landing procedures are critical.

The criticality of the aforementioned applications imposes an error tolerant constraint on systems. Furthermore, space is a harsh radiation environment, which negatively affects electronic devices. Errors produced by radiation in electronic devices are mainly categorised into two types: hard and soft errors. Hard errors are permanent changes to the device or circuit, while soft errors are non-destructive errors that corrupt stored information. They are typically Single Event Upsets (SEU) or Multiple Bit Upsets (MBU), with the main difference between the two being how many logic elements are corrupted. 

Developing radiation-hardened hardware has its disadvantages, such as an increase of production costs and that technology usually is some generations behind modern ones. This disadvantages make designers to shift into Commercial-Off-The-Shelf (COTS) devices, which offer great performance at a fair cost and to aim at soft error mitigation techniques. To mitigate soft errors, Radiation Hardening By Design (RHBD) techniques are typically used, with some examples being Error Correction Codes (ECC) for memories to Dual Modular Redundancy (DMR) and Triple Modular Redundancy (TMR) for block designs.

Rather than using conventional hardware redundancy techniques such as DMR or TMR, we exploit the class relationships that CNNs create when learning. As these relationships are not explicit, we use an ANN to learn them and infer when they are broken. This way, misclassification errors produced by soft errors in parameters are detected. In this work, we propose a solution for SEU mitigation in CNNs called Error Supervising Neural Network (ESupNNet). This architecture combines the main CNN that needs protection or target CNN, with an auxiliary ANN used to detect errors. The main advantages of ESupNNet are its low overhead, low training cost and non-modification of the target CNN. The proposal is validated with two widely used imagery datasets: ImageNet-1K \cite{imagenet1k} and CIFAR-10 \cite{cifar10}, and for five different CNN architectures: ResNet-50 \cite{resnet50}, EfficientNetV2-Small \cite{efficientnetv2}, MobileNetV3-Small \cite{mobilenetv3}, ShuffleNetV2 with 2.0x output channels \cite{shufflenetv2} and MNASNet with depth multiplier of 1.3 \cite{mnasnet}.

The rest of the paper is organised as follows: Section II shows a revision of the literature regarding fault tolerance in ANNs; Section III describes the architecture proposed; Section IV details the experiments done to validate the solution, together with the results obtained and a comparison with existing techniques; finally, Section V shows the conclusions of this work.

\section{Related Work}
\label{sec:relatedwork}
Fault tolerance of ANNs is widely studied because they are affected by imprecision, uncertainty and faults \cite{ft_0, ft_1}. A typically chosen device to implement ANNs is the Field Programmable Gate Array (FPGA). They offer great computing parallelisation, which can be exploited by ANNs to obtain maximum throughput and low power consumption, which is intended for embedded systems. Reliability analyses are being done for such implementations \cite{ft_2,ft_3}, which mainly explore how different quantisation levels, architectures, and resources affect ANNs when altered by radiation-induced faults.

\subsection{Hardware redundancy techniques}
In digital computing systems, reliability against soft errors is achieved with redundancy information or time. It is important to note that improving the reliability of a system is a trade-off between fault tolerance, resources and performance. Typically, Triple Modular Redundancy (TMR) is used with a voting mechanism. Theoretically, TMR gives the system full fault tolerance at the expense of power consumption and more than x2 resource overhead. Therefore, a more selective approach to redundancy is to reduce the increase of resources while achieving acceptable fault tolerance capabilities. 

In \cite{stmr_0}, selective TMR is used to triplicate critical layers of a CNN while reducing the resource overhead of a full TMR design. The authors analyse the reliability of two CNNs with fault injection to obtain which layers are more critical and apply TMR only to those layers. The advantage of selective TMR in CNNs is that not all errors produce a wrong inference, as forward propagation can mitigate faults and the classification layer at the end can mask some output deviations \cite{cnnresilience_0,cnnresilience_1}. Thus, instead of protecting against all errors, it is more relevant to mitigate errors that produce a wrong classification. In \cite{atmr_0}, authors use approximate TMR in critical elements to reduce the resource overhead of selective TMR at the expense of less error masking performance. With approximate TMR applied to specific layers, the overhead of each TMR layer is reduced as the complexity of this TMR is lower by means of approximate computing. Another form of reducing resource overhead of selective TMR, is to apply TMR to critical elements and then eliminate redundant parameters of the CNN like in \cite{stmr_1}. Performing a thorough analysis of the ANN in order to determine which layers are more critical is not trivial. First, the fault injection campaign needs to be statistically accurate in order to correctly detect critical layers. Then, those layers need to be protected with TMR or approximate TMR, and a new fault injection campaign is required to validate the solution.

On the other hand, studies like \cite{watk_0} focus on the protection of parameters like weights against attacks. This study can be extrapolated to protect parameters of ANNs against radiation-induced soft errors. In their work, authors propose an autoencoder to compress layers and store them redundant in a latent representation. As information is compressed, storing data in a redundant manner incurs minimal overhead. When performing inferences, the compressed layers are voted to correct errors and reconstructed by a decoder. This solution effectively mitigates errors in weights, but the reconstructed model suffers a degradation in accuracy.

\subsection{Fault-aware training}
Hardware redundancy techniques are not always suitable for implementation, as these solutions are typically used for FPGA based designs that easily allow its integration. For other devices such as Graphics Processing Units (GPUs), running these mechanisms is not straightforward and can impose a reduction in performance due to the voting or comparison mechanism. One solution to increase the error tolerance of ANNs is to perform fault-aware training. With this method, soft errors are injected during training to make the ANN resilient to errors. The network learns to adjust its parameters when affected by errors to avoid degradation of its accuracy. In \cite{retrain_0,retrain_1,retrain_2,retrain_3} fault-aware training is used to improve the reliability of CNNs against soft errors. Although this method increases the resilience of CNNs, there are some problems. First, the whole CNN needs to be trained taking into account errors, which can take a toll on its accuracy regarding the absence of errors. Secondly, training CNNs of considerable size consumes a great amount of energy. If more data with faults are needed, the training process needs more computations, thus, more energy consumption. Lastly, fault-aware training does not allow to directly take an already trained and validated CNN to be deployed in an error-prone environment. When using CNNs of considerable size, fault-aware training might incur in significant training costs.

\subsection{Algorithm-based techniques}
Other methods to increase computing reliability include Algorithm-Based Fault Tolerance (ABFT) and Algorithm-Based Error Detection (ABED)\cite{abft_0}, which consists on encoding data, apply an algorithm in the encoded data and output encoded data. This technique allows error detection and correction directly in the data structure, instead of relying on the hardware. In \cite{abft_1,abft_2,abft_3} authors apply ABFT and ABED techniques to increase the reliability of convolutional operations in CNNs. These methods offer fault coverage for convolutional operations, but are not suitable to mitigate soft errors affecting the parameters of ANNs as protection is focused on the operations.

\subsection{Summary}
One of the problems of fault tolerance mechanisms is that the circuit, model or process to be protected needs to undergo some kind of modification. In redundancy, the protected model needs to be replicated either completely or partially to detect or correct errors (depending on the replicas) with the subsequent overhead. In fault-aware training, the protected model needs to undergo new training and validation phases with a modified dataset to be able to withstand faults once deployed. This process can be very time- and energy-costly for state-of-the-art or complex ANNs. In ABFT, the protected model needs to change the way operations are done to mitigate errors during execution and does not necessarily protect against soft errors affecting parameters. To reduce the resource- and energy- cost requirements of these solutions, it is important to focus on what can be done without altering existing models.

\section{Error Supervising Neural Network architecture}

\subsection{Error detection}

Modern CNNs are typically made of two parts, the feature extractor and the classifier, each with their own set of parameters: $\theta_f$ for the feature extractor and $\theta_c$ for the classifier. Let $\Phi_{\theta_f}: X \rightarrow \mathbb{R}^m$ denote the feature extractor learnt by a trained CNN. Using filters or kernels, the feature extractor creates a feature space. After this mapping, every input sample $x$ becomes a point in a m-dimensional space from which class relations can be obtained. As an example, each class $y$ can be represented by its centroid in this m-dimensional space and pairwise similarities between these centroids define a class relation matrix.

\begin{gather}
	\mu_i = \frac{1}{N_i}\sum_{x:y=i}\Phi_{\theta_f}(x)
	\\
	R_{ij} = s(\mu_i,\mu_j)
\end{gather}
where $N_i$ is the number of samples in class $i$ and $s(\cdot,\cdot)$ is a similarity measure.

The fully connected (FC) part transforms the geometry of the feature space to improve classification. As the FC part reshapes the geometry, semantic relations are transformed during which some are preserved and others are compressed. The implicit class relations are encoded in the composition of all FC layers $F_{\theta_c} = f_1 \circ f_2 \circ \ldots \circ f_L$, which typically is a non-linear mapping. Finally, the output logits contain information about the relations between classes inherited from the feature space. As an example, each class can be represented by its mean logit vector and the similarities between these vectors define a class relation matrix.

\begin{gather}
	z(x) = F_{\theta_c}(\Phi_{\theta_f}(x))
	\\
	R_{ij} = s(E[z(x) \mid y=i], E[z(x) \mid y=j])
\end{gather}
where $s(\cdot,\cdot)$ is a similarity measure.

The main idea of this work is to obtain a class relation $R$ such that:
\begin{gather}
	\hat{y} = classification(z(x))
	\\
	z(x)\in R \iff \hat{y} = y
\end{gather}
With $y$ being the ground truth and $\hat{y}$ the predicted class.

Therefore, classification errors can be detected if the output logits do not belong to the class relation. To illustrate this idea, we designed a simple CNN for the MNIST dataset \cite{mnist} from torchvision library \cite{torchvision2016} using Pytorch framework \cite{pytorch}. The CNN consists of three convolutional layers, each with ReLU as activation function followed by a pooling layer; two FC layers, with the first one having ReLU as activation function; and a final softmax layer to obtain the probability of classes. The loss function is Cross-Entropy and Stochastic Gradient Descent (SGD) the optimisation function. We used a learning rate of 0.0001 and a momentum of 0.5, both constant for 100 epochs. The distribution of the dataset for training and validation is 75\%/25\% with a total 70,000 images and training data is randomly shuffled each epoch. The resulting CNN has an accuracy of 99.04\%.

\begin{figure}
	\centering
	\includegraphics[width=\columnwidth]{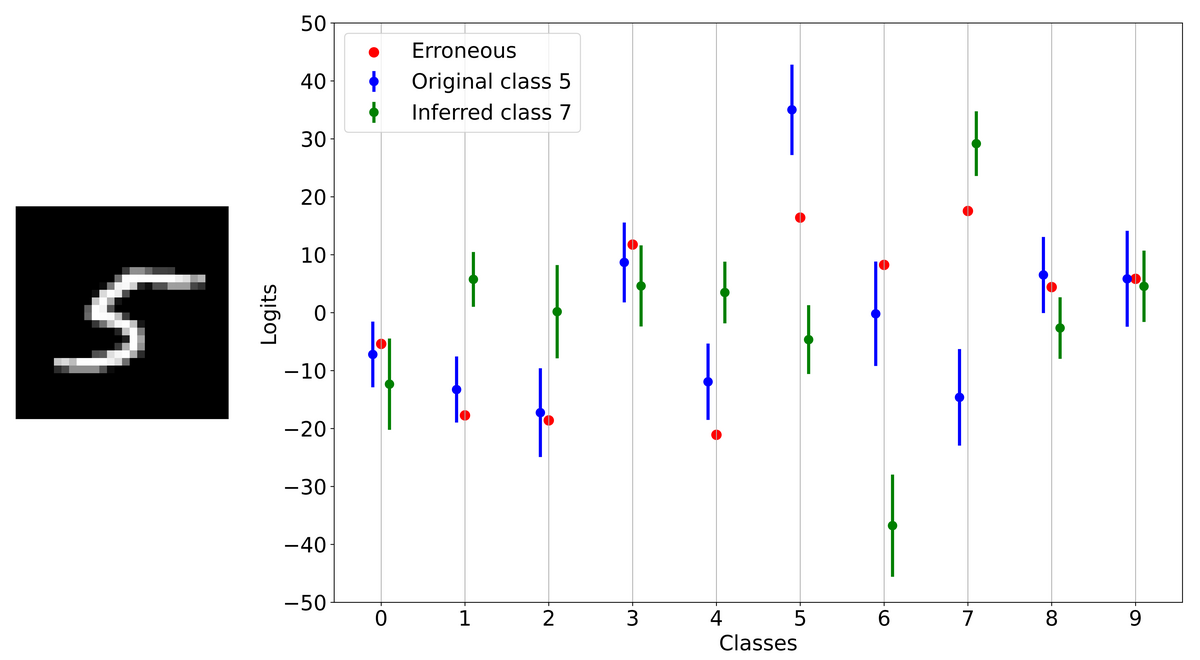}
	\caption{Example of the effects in class relations of a SEU in the parameters of a CNN. On the left is the sample image used, on the right the value of output logits of the CNN during the erroneous inference and the error around the mean value of logits for the original and inferred classes.}
	\label{fig:example}
\end{figure}

We performed a random bitflip in a random parameter of the CNN, simulating an SEU. Then, used the CNN with the bitflip to infer an image from the validation dataset. Fig. \ref{fig:example} shows the results of this small experiment. On the left is the sample image, which corresponds to the class of handwritten fives. On the right there is a graph showing the value of the output logits of the CNN. The graph shows the mean value and standard deviation of each output logit for the ground truth and inferred classes (handwritten five and seven respectively) using the original CNN over the training dataset. These are plotted together with the value of the output logits of the altered CNN. Although no explicit relation between classes is calculated, a simple graphical relation can be obtained. For samples of the class ``5'' (blue in Fig. \ref{fig:example}), classes ``3'', ``8'' and ``9'' have very similar values being the highest after ``5'', and class ``7'' is among the lowest. Whether for samples of the class ``7'' (green in Fig. \ref{fig:example}), ``1'', ``3'' and ``9'' are among the highest after ``7'', and class ``6'' is the lowest. These results show that the CNN has established, through its parameters, a relation between classes that helps to classify them based on the features extracted by $\Phi_{\theta_f}$. The logits obtained in the erroneous inference (red in Fig. \ref{fig:example}) have a relationship that is very different to the one expected for class ``7''. For example, the expected logit for ``6'' is the lowest, while the one obtained is among the highest (among other differences). On the other hand, the erroneous inference shows a similar relation between classes than the calculated for class ``5''. The obtained relation is very different to the one calculated for inferences belonging to class ``7'', therefore the error can be detected.

Modern CNNs are complex systems and the broad spectrum of errors in parameters make obtaining $R$ a complicated problem. The induced class relations heavily depend on the trainable model and the separability of classes. Thus, it is a good strategy to use an ANN to obtain an approximate relation $R^*$. It might also be possible to use these inter-class relationships not only to detect, but to correct misclassification errors. On this matter, the ANN used to obtain $R^*$ would learn to perform inferences in an environment prone to errors.

\subsection{Architecture}

\begin{figure*}
	\centering
	\setlength{\fboxrule}{0pt}
	\fbox{\includegraphics[width=0.8\textwidth]{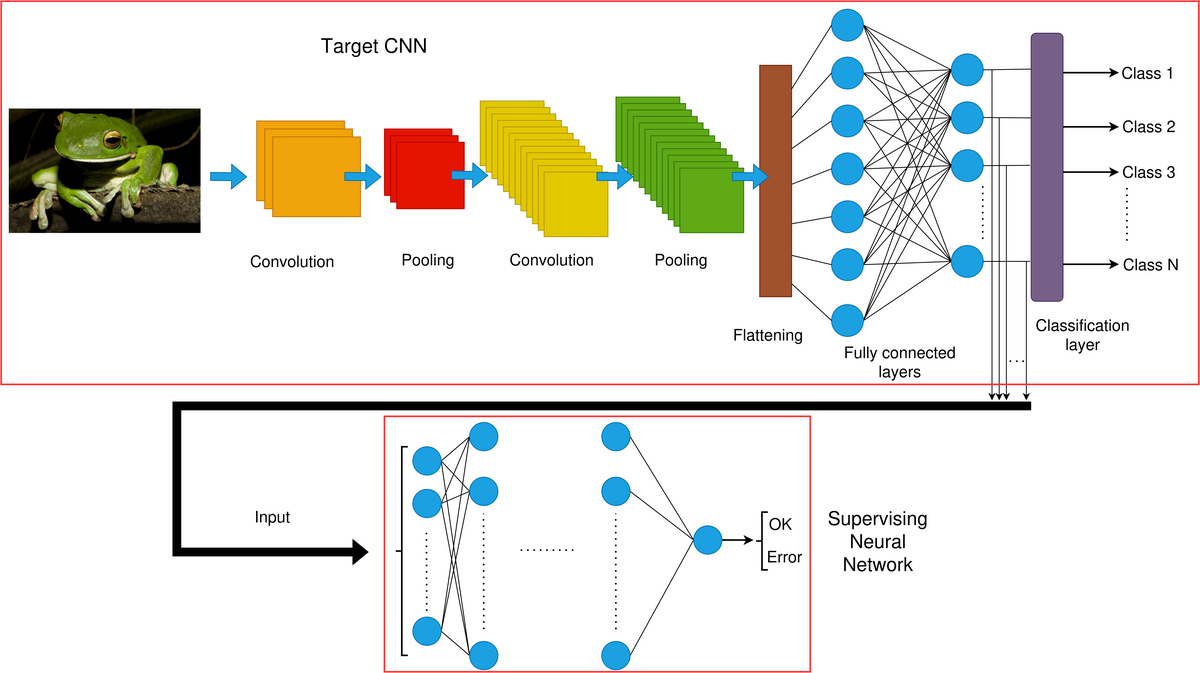}}
	\caption{Example of the proposed ESupNNet architecture for error detection. Sample image taken from ImageNet-1K dataset \cite{imagenet1k}.}
	\label{fig:arch}
\end{figure*}

We propose a low overhead, non-intrusive Error Supervising Neural Network architecture for fault tolerance against soft errors of parameters called ESupNNet. This architecture can be seen in Fig. \ref{fig:arch}, which consists of a CNN that needs to be protected (hereinafter target CNN) and a supervising ANN. No alterations of the target CNN like redundancy, modification of operations or retraining are required. The supervising ANN is a binary classifier that takes as inputs the output logits of the target CNN last layer (before classification layer) to detect if there are any errors in its classification. The proposed architecture can include not only the output logits of the target CNN, but the output of other internal neurons. We decided to use only the output logits because all the information inside the target CNN is compressed to these elements. Thus, it reduces the complexity of training the supervising ANN and generalises this solution because the target CNN can be treated as a black-box model.

In a general way, the architecture can be expressed as in Eq. \ref{eq_0}. 

\begin{equation}
	\begin{split}
		&M:X^n \rightarrow Y^m \xrightarrow{classification} \{0, ..., m-1\}
		\\
		&S:Y^m \rightarrow \mathbb{R} \xrightarrow{\sigma} [0, 1] \xrightarrow{threshold} \{0, 1\}
	\end{split}	
	\label{eq_0}
\end{equation}
With $M$ being the target CNN, $S$ the supervising ANN, $X$ the input of the target CNN, $Y$ the output of the target CNN and $n$, $m$ the dimensions of the input and output respectively.

To detect classification errors due to parameter alteration, the supervising ANN implements the following expression
\begin{gather}
	\intertext{Be $R^*$ the approximate relation to $R$}
	\text{if}\ z(x) \notin R^* \Rightarrow \hat{y} \neq y
\end{gather}

If $\theta'$ is the set of parameters altered by an error. The resulting logits are
\begin{gather}
	z'(x) = F_{\theta_c}(\Phi_{\theta_f}(x)) \ | \ \theta_c \in \theta' \vee \theta_f \in \theta'
	\\
	\hat{y}' = classification(z'(x))
\end{gather}

$R^*$ is calculated by regression, using a dataset $D$ of elements containing the output logits of inferences as input data for the supervising ANN and if the classification is correct as label. Using Kronecker delta ($\delta$), the dataset is expressed as
\begin{equation}
\begin{split}
	D = (z(x), \delta_{\hat{y}y} = 1)\ &\cup \ (z'(x), \delta_{\hat{y}'y} = 1) \\ &\cup \ (z'(x), \delta_{\hat{y}'y} = 0) \\
	\forall x \ | \ \delta_{\hat{y}y} = 1
\end{split} 
\label{eq_dataset}
\end{equation}

By Eq. \ref{eq_dataset}, the dataset contains the output logits of the unaltered model and of the model with altered parameters. It shall be noted that all inferences, both for models with and without altered parameters, are done for data that is classified correctly by the unaltered model. By calculating $R^*$ with regression for dataset $D$ using an ANN, the model is able to take into account that if $z'(x)$ does not produce a wrong classification ($\delta_{\hat{y}'y} = 1$), it belongs to $R^*$. To obtain the output logits of the altered model, the model parameters need to be altered first and the input data is processed by the model for inference. This can be done by simulation, which makes obtaining the dataset a feasible process.

The ESupNNet architecture provides several advantages compared to solutions reviewed in Sec. \ref{sec:relatedwork}. The first one is that this architecture is non-intrusive. No modifications of the target CNN are needed such as redundancy, retraining or modification of operation. Secondly, the supervising ANN does not incur in a relevant overhead as its input layer has the same dimensionality of the output layer of the CNN. The output layer of the supervising ANN consists of one artificial neuron, so an encoder scheme can be used. Thirdly, generating dataset $D$ can be done by simulation, which allows to easily obtain a significant amount of data without the need of external input like new real world images of the classes.

\section{Experimental Results}

This section is organised as follows: 
\begin{itemize}
	\item First, we explain which benchmarks such as datasets and CNNs are used to validate the proposed architecture.
	\item Then, we show how the dataset used to calculate $R^*$ is obtained.
	\item After that, the architecture of the supervising ANN is described.
	\item Lastly, we show how the training and validation of ESupNNet is done, together with the results obtained.
\end{itemize}

The experimental platform used in this project is a desktop computer with AMD Ryzen 5 3600X processor and NVIDIA GeForce GTX 1660 (6G) graphics card. The operating system is Ubuntu 24.04.4 LTS. The development language is Python 3.12.2. All benchmark models and datasets used were obtained through torchvision \cite{torchvision2016} and the deep learning framework is Pytorch \cite{pytorch}.

\subsection{Selected benchmarks}
To test the performance of ESupNNet, we selected two imagery datasets: ImageNet-1K \cite{imagenet1k} and CIFAR-10 \cite{cifar10}. This two datasets are chosen because of their different complexity and to the different number of classes (1,000 vs 10). The relation $R^*$ is obtained using the output logits, whose dimension is the same that the number of classes. Therefore, the amount of information used to obtain $R^*$ is different for both datasets and can have an impact on the approximation of $R^*$ to $R$. To validate the proposed architecture, five CNN architectures were selected: ResNet-50 \cite{resnet50}, EfficientNetV2-Small \cite{efficientnetv2}, MobileNetV3-Small \cite{mobilenetv3}, ShuffleNetV2 with 2.0x output channels \cite{shufflenetv2} and MNASNet with depth multiplier of 1.3  \cite{mnasnet}. We made this selection of models because they vary in number of parameters, which is relevant to test if the proposed architecture can be used for different devices like GPUs and FPGAs. Also, we can analyse the performance of ESupNNet across different architectures. Each architecture has its own feature extractor and classifier, so their feature space and inherited class relations are different. 

Models ResNet-50 and EfficientNetV2-Small are used with ImageNet-1K, mainly because of the large number of parameters. No retraining is needed as the parameters from torchvision are trained for this dataset. Models MobileNetV3-Small, ShuffleNetV2 and MNASNet are used with CIFAR-10, mainly because they have less parameters than the models before and can be possible candidates for embedded systems. As the parameters from torchvision are trained for ImageNet-1K, some changes are needed to be used for CIFAR-10. We changed the output layer from 1,000 neurons to 10 neurons to meet CIFAR-10 classes and retrained the target CNNs for 10 epochs. The distribution of the training and validation datasets is 83\%/17\% (taken from torchvision datasets), with training data being randomly shuffled each epoch. The loss function is Cross-Entropy and the optimisation function SGD. We used a constant learning rate of 0.001 and a momentum of 0.9. The final Top-1 accuracy of each model is given in Table \ref{tab:accnominal}, with an accuracy over 90\% for models using CIFAR-10.

\begin{table}[]
	\centering
	\caption{Models used as benchmarks and their accuracy to respective datasets}
	\label{tab:accnominal}
	\begin{tabular}{lcc}
		\hline
		\multicolumn{1}{c}{\textbf{Model}} & \textbf{Dataset} & \textbf{Accuracy} \\ \hline
		ResNet-50                   & ImageNet-1K & 80.86\%  \\
		EfficientNetV2-Small       & ImageNet-1K & 84.23\%  \\ \hline
		MobileNetV3-Small          & CIFAR-10 & 93.21\%  \\
		ShuffleNetV2               & CIFAR-10 & 95.12\%  \\
		MNASNet                    & CIFAR-10 & 91.68\%  \\ \hline
	\end{tabular}
\end{table}

\subsection{Datasets}
As introduced in Eq. \ref{eq_dataset}, the dataset $D$ used to train the supervising ANN needs three different types of data regarding the target CNN: data correctly classified without model alteration (correct $X_c$), data correctly classified with model alteration (silent $X_s$) and data wrongly classified with model alteration (errors $X_e$). These data must be obtained using inputs to the target CNN that are inferred correctly without model alteration to extract its class relations.

The dataset for each model is different, but they all follow the same distribution of data in a balanced way. A detailed summary is given in Table \ref{tab:dataset}. $X_c$ and $X_s$ are data that do not produce a wrong inference, while $X_e$ are logits from wrong inferences. Thus, to balance the dataset, we decided that 50\% of the dataset is made of $X_e$, while the other half is divided in two equal parts for $X_c$ and $X_s$. On the other hand, the set of parameters of the target CNN are weights and bias, with weights containing kernel values and neuron weights. The number of parameters that are bias is lower than weights, ranging from 1\% to 5\%. An error in a weight affects the input it uses and its effect can be mitigated by the neuron itself. An error in a bias affects the whole neuron, so they are more dangerous. To avoid creating a huge dataset and make errors in bias relevant, we decided to distribute data of altered parameters in a ratio of 90\%/10\% for weights ($X_{*w}$) and bias ($X_{*b}$) respectively. Furthermore, data from all subsets is equally distributed for each class so ensure balance. As an example from Table \ref{tab:dataset}, in CIFAR-10 where there are 10 classes, each class in the subset $X_c$ would correspond to a 2.5\% of the total dataset. Taking $X_{eb}$, each class would correspond to a 0.5\% of the total dataset.

\begin{table*}[]
	\centering
	\caption{Distribution of all types of data in dataset used by supervising ANN}
	\label{tab:dataset}
	\begin{tabular}{lcccccc}
		\cline{4-7}
		& \multicolumn{1}{l}{}                                   & \multicolumn{1}{l}{}                                 & \multicolumn{2}{c}{\textbf{$\mathbf{X_s}$ (25\%)}}                                                                     & \multicolumn{2}{c}{\textbf{$\mathbf{X_e}$ (50\%)}}                                                                    \\ \cline{2-7} 
		& \textbf{Total}                                         & \textbf{$\mathbf{X_c}$ (25\%)}                                   & \textbf{$\mathbf{X_{sw}}$ (22.5\%)}                                & \textbf{$\mathbf{X_{sb}}$ (2.5\%)}                                & \textbf{$\mathbf{X_{ew}}$ (45\%)}                                  & \textbf{$\mathbf{X_{eb}}$ (5\%)}                                 \\ \hline
		\textbf{ResNet-50}         & \begin{tabular}[c]{@{}c@{}}800,000\\ 800\end{tabular}   & \begin{tabular}[c]{@{}c@{}}200,000\\ 200\end{tabular} & \begin{tabular}[c]{@{}c@{}}180,000\\ 180\end{tabular} & \begin{tabular}[c]{@{}c@{}}20,000\\ 20\end{tabular}  & \begin{tabular}[c]{@{}c@{}}360,000\\ 360\end{tabular} & \begin{tabular}[c]{@{}c@{}}40,000\\ 40\end{tabular} \\ \hline
		\textbf{EfficientNetV2-S} & \begin{tabular}[c]{@{}c@{}}800,000\\ 800\end{tabular}   & \begin{tabular}[c]{@{}c@{}}200,000\\ 200\end{tabular} & \begin{tabular}[c]{@{}c@{}}180,000\\ 180\end{tabular} & \begin{tabular}[c]{@{}c@{}}20,000\\ 20\end{tabular}  & \begin{tabular}[c]{@{}c@{}}360,000\\ 360\end{tabular} & \begin{tabular}[c]{@{}c@{}}40,000\\ 40\end{tabular} \\ \hline
		\textbf{MobileNetV3-S}    & \begin{tabular}[c]{@{}c@{}}172,000\\ 17,200\end{tabular} & \begin{tabular}[c]{@{}c@{}}43,000\\ 4,300\end{tabular}  & \begin{tabular}[c]{@{}c@{}}38,700\\ 3,870\end{tabular} & \begin{tabular}[c]{@{}c@{}}4,300\\ 430\end{tabular} & \begin{tabular}[c]{@{}c@{}}77,400\\ 7,740\end{tabular} & \begin{tabular}[c]{@{}c@{}}8,600\\ 860\end{tabular} \\ \hline
		\textbf{ShuffleNetV2}     & \begin{tabular}[c]{@{}c@{}}172,000\\ 17,200\end{tabular} & \begin{tabular}[c]{@{}c@{}}43,000\\ 4,300\end{tabular}  & \begin{tabular}[c]{@{}c@{}}38,700\\ 3,870\end{tabular} & \begin{tabular}[c]{@{}c@{}}4,300\\ 430\end{tabular} & \begin{tabular}[c]{@{}c@{}}77,400\\ 7,740\end{tabular} & \begin{tabular}[c]{@{}c@{}}8,600\\ 860\end{tabular} \\ \hline
		\textbf{MNASNet}          & \begin{tabular}[c]{@{}c@{}}172,000\\ 17,200\end{tabular} & \begin{tabular}[c]{@{}c@{}}43,000\\ 4,300\end{tabular}  & \begin{tabular}[c]{@{}c@{}}38,700\\ 3,870\end{tabular} & \begin{tabular}[c]{@{}c@{}}4,300\\ 430\end{tabular} & \begin{tabular}[c]{@{}c@{}}77,400\\ 7,740\end{tabular} & \begin{tabular}[c]{@{}c@{}}8,600\\ 860\end{tabular} \\ \hline
	\end{tabular}
\end{table*}

To obtain the dataset, we performed fault injection in the parameters of the target CNN before doing an inference. We made a custom fault injector program to produce bitflips in the parameters of the target CNNs. To do the fault injection, a random parameter is chosen, transformed to signed fixed point format Q6.10, a bitflip is done in a random bit and transformed back to its original format. This bit width was chosen as all parameters could be represented with it and 16 bits is a common way of compressing the storage of parameters for memory constrained and embedded systems without losing much information. Then, the altered target CNN is given an input data from the original dataset. Depending on the classification result, it is added to $X_s$ if silent or $X_e$ if it produced an error. This process was done for each model and the respective classes of its dataset. To comply with the balance of the dataset, the number of datapoints of each class that are originally correctly classified ($X_c$) is obtained. This number is used as the upper limit of the number of elements each class has in $D$. The size of $D$ and its respective types of data are reported in Table \ref{tab:dataset}. In each cell, the upper value corresponds to the total number of elements of the data type denoted by the column. The lower value corresponds to the total number of elements of each class for such type. No data augmentation is done to the original datasets in order to obtain $D$.

The large number of parameters, the bit width and the number of original data per class make necessary to perform sampling through fault injection. The total population is equal to the Cartesian product of these sets. A trade-off between dataset size, time needed to obtain data and statistical significance limits the generation of the dataset. During experiments, $D$ is a subset of another parent dataset $D_p$, which was obtained by sampling and is not completely balanced. Table \ref{tab:errormargins} shows the mean number of fault injections done for each class and parameter type. It also shows the error margin ($e$) obtained using Cochran formula \cite{cochran1977}, shown in Eq. \ref{eq_cochran}. 

\begin{equation}
n_0 = \frac{Z^2p(1-p)}{e^2}
\label{eq_cochran}
\end{equation}

To calculate the error margin, a 99\% confidence interval ($Z$=2.576) and $p$=0.5 are used, with $n_0$ being the number of injections performed. It is observed that the error margin for models that use ImageNet-1K dataset is bigger than for those that use CIFAR-10. This is because the process to obtain fault injection data for $D_p$ was more demanding for the former ones, given the experimental platform available, so we reduced the total number of injections. Nonetheless, it is important to mention that in a 99\% confidence interval, the error margin is less than 2\% in the worst case (injections in biases for ResNet-50), and less than 0.4\% for those using CIFAR-10. These results provide a strong statistical relevance to ensure that $D_p$ is a correct representation of the effects of SEUs from which to obtain $R^*$.

\begin{table}[]
	\centering
	\caption{Error margin of the parent dataset $D_p$ calculated using Cochran formula with Z=2.576 and p=0.5 for the injections done per class}
	\label{tab:errormargins}
	\begin{tabular}{lrrrr}
		\cline{2-5}
		& \multicolumn{2}{c}{\textbf{\begin{tabular}[c]{@{}c@{}}Injections per class\end{tabular}}} & \multicolumn{2}{c}{\textbf{Error margin (\%)}}                          \\ \cline{2-5} 
		& \multicolumn{1}{c}{\textbf{weight}}              & \multicolumn{1}{c}{\textbf{bias}}             & \multicolumn{1}{c}{\textbf{weight}} & \multicolumn{1}{c}{\textbf{bias}} \\ \hline
		\textbf{ResNet-50}         & 21,694                                            & 4,172                                          & 0.87                                & 1.99                              \\ \hline
		\textbf{EfficientNetV2-S} & 17,832                                             & 12,093                                          & 0.96                                & 1.17                              \\ \hline
		\textbf{MobileNetV3-S}    & 825,000                                           & 138,000                                        & 0.14                                & 0.35                               \\ \hline
		\textbf{ShuffleNetV2}     & 703,000                                           & 218,000                                        & 0.15                                & 0.28                              \\ \hline
		\textbf{MNASNet}          & 189,000                                           & 121,000                                        & 0.3                                 & 0.37                              \\ \hline
	\end{tabular}
\end{table}

The structure of the parent dataset $D_p$ from the total fault injection campaign is reported in Table \ref{tab:parentd}. The table shows the number of elements per class regarding type of data. For the column $X_c$ the mean value (top) and the range (down) are shown because it corresponds to the original dataset and the number of elements per class that are correctly classified is not balanced. For the following columns, the number of elements per class is balanced because they were obtained through fault injection. The same number of elements were generated for the subtype (weight or bias) independently of being silent ($X_{s*}$) or causing an error ($X_{e*}$).

\begin{table}[]
	\centering
	\caption{Number of elements per class in parent datasets $D_p$}
	\label{tab:parentd}
	\begin{tabular}{lrrr}
		\cline{2-4}
		& \multicolumn{1}{c}{$\mathbf{X_c}$}                                 & \multicolumn{1}{c}{$\mathbf{X_{*w}}$} & \multicolumn{1}{c}{$\mathbf{X_{*b}}$} \\ \hline
		\textbf{ResNet-50}         & \begin{tabular}[c]{@{}c@{}}1,215\\ {[}582, 1350{]}\end{tabular}  & 400                              & 200                              \\ \hline
		\textbf{EfficientNetV2-S} & \begin{tabular}[c]{@{}c@{}}1,245\\ {[}503, 1350{]}\end{tabular}  & 400                              & 200                              \\ \hline
		\textbf{MobileNetV3-S}    & \begin{tabular}[c]{@{}c@{}}5,762\\ {[}5511, 5894{]}\end{tabular} & 10,000                            & 5,000                             \\ \hline
		\textbf{ShuffleNetV2}     & \begin{tabular}[c]{@{}c@{}}5,922\\ {[}5839, 5972{]}\end{tabular} & 10,000                            & 5,000                             \\ \hline
		\textbf{MNASNet}          & \begin{tabular}[c]{@{}c@{}}5,689\\ {[}5158, 5880{]}\end{tabular} & 10,000                            & 5,000                             \\ \hline
	\end{tabular}
\end{table}

\subsection{Supervising ANN}
The supervising ANN architecture was designed to have a small footprint while maintaining a good performance. Although each benchmark model is different, we wanted to show that the ESupNNet architecture can be used in a general way for CNN models. Thus, the same architecture is used for those CNN models using the same dataset, as the dimensionality of the input layer is the same as that of the output logits of the target CNN (number of classes of original dataset). Following Fig. \ref{fig:arch} and Eq. \ref{eq_0}, the architecture of the supervising ANN is a high dimensionality input to one dimension output (error, no error). The input of the target CNN is normalised by Min-Max scaling to make the supervising ANN extract class relations without relying only on abnormal values. A FC ANN is used with LeakyReLU as activation function for all neurons except the output one. For classification, the sigmoid function with a threshold at 0.5 is applied to the output logit. The number of neurons per layer from input to output is:

\begin{itemize}
	\item For ResNet-50, EfficientNetV2-Small: \\1,000-250-100-1.
	\item For MobileNetV3-Small, ShuffleNetV2, MNASNet: \\10-100-80-60-40-20-1.
\end{itemize}

It is seen that the number of layers is bigger for models using CIFAR-10 than for ImageNet-1K. For models using ImageNet-1K, the number of neurons of following layers is strictly lower. For those using CIFAR-10, the number of neurons in the second layer is increased before reducing it in following layers. 

To obtain $R^*$, the information carried by the output logits of the target CNN has to be relevant enough to extract meaningful class relations. Thus, with an input of low dimensionality, it is better to first expand information before starting to compress it. For models using CIFAR-10, the input of the supervising ANN is 10-dimensional, so we increased the number of neurons in the second layer before starting to compress it. This idea was tested experimentally and for models using CIFAR-10, the structure shown proved to have good performance while incurring in low resources overhead and training times. For models using ImageNet-1K, which has 1,000 classes, it was not necessary.

The parameters overhead of the supervising ANN compared to its target CNN is the following:

\begin{itemize}
	\item ResNet-50: 0.86\%
	\item EfficientNetV2-Small: 1.03\%
	\item MobileNetV3-Small: 1.13\%
	\item ShuffleNetV2 x2\_0: 0.32\%
	\item MNASNet 1\_3: 0.35\%
\end{itemize}

To perform a fair analysis, the training process was decided to be the same for all models. K-folds was used for cross-validation, with K=10. Each dataset $D$ was randomly obtained from its respective parent dataset $D_p$, following the balanced structure detailed in Table \ref{tab:dataset}. The distribution of $D$ for training and validation is 80\%/20\% respectively, with a random shuffle of the training set each epoch. The loss function chosen is Binary Cross-Entropy (BCE) Loss, implemented in Pytorch as the function BCEWithLogitsLoss due to its numerical stability compared to a sigmoid layer followed by BCE Loss. The optimiser is SGD with a learning rate of 0.01 and a momentum of 0.9. These values are not constant, for which an scheduler was used taking into account validation loss. The scheduler was implemented using Pytorch function ReduceLROnPlateau. The learning rate factor is 0.1, with a patience of 5 epochs, a threshold of 0.01 and a cooldown of one epoch. The training process is done for 50 epochs. The batch size is 512 for target CNNs using ImageNet-1K dataset and 64 for models using CIFAR-10 dataset.

\subsection{Results}

\begin{figure*}[]
	\captionsetup[subfigure]{labelformat=empty}
	\centering
	\subfloat[]{
		\includegraphics[width=0.4\textwidth]{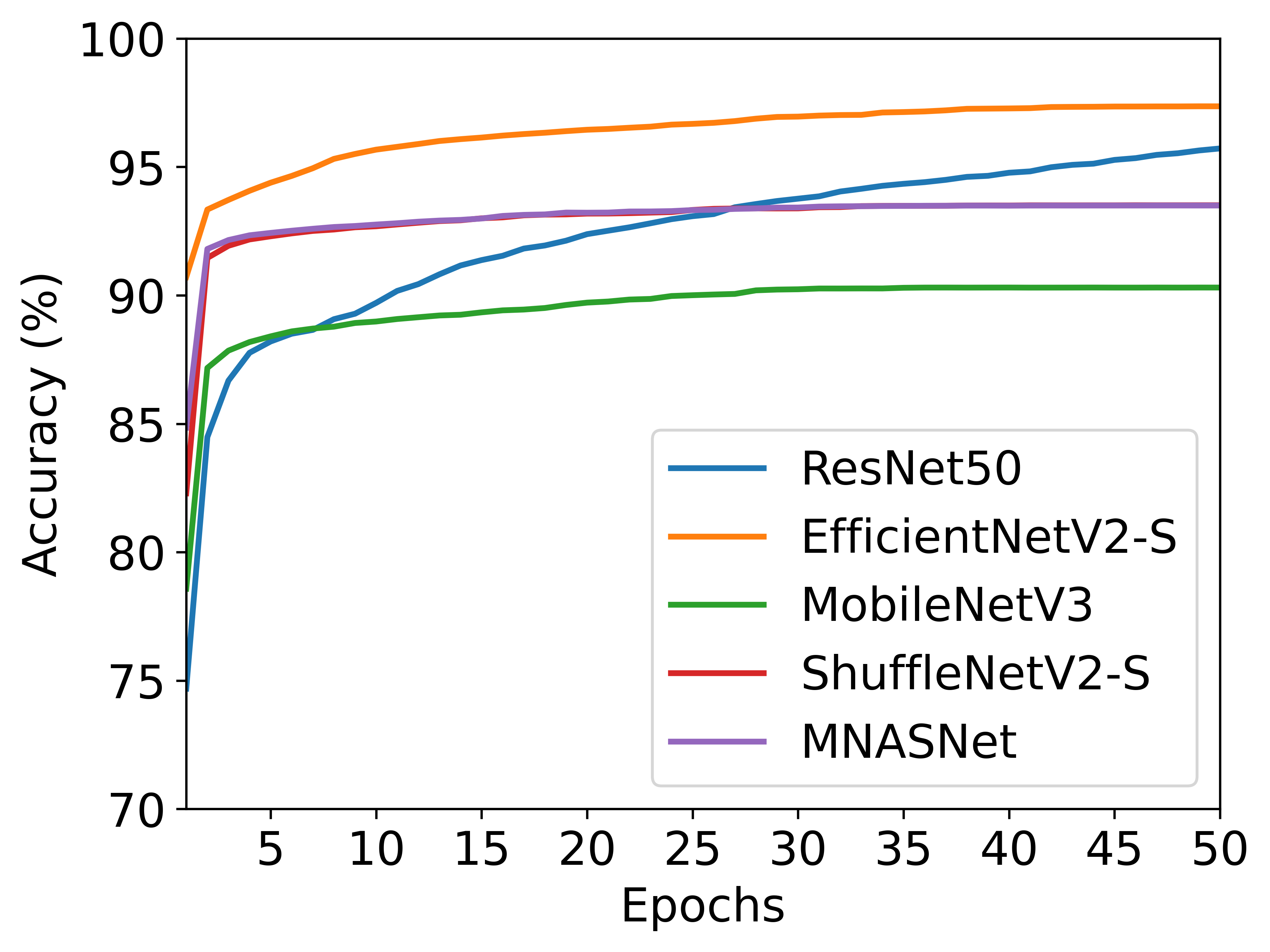}
	}
	\subfloat[]{
		\includegraphics[width=0.4\textwidth]{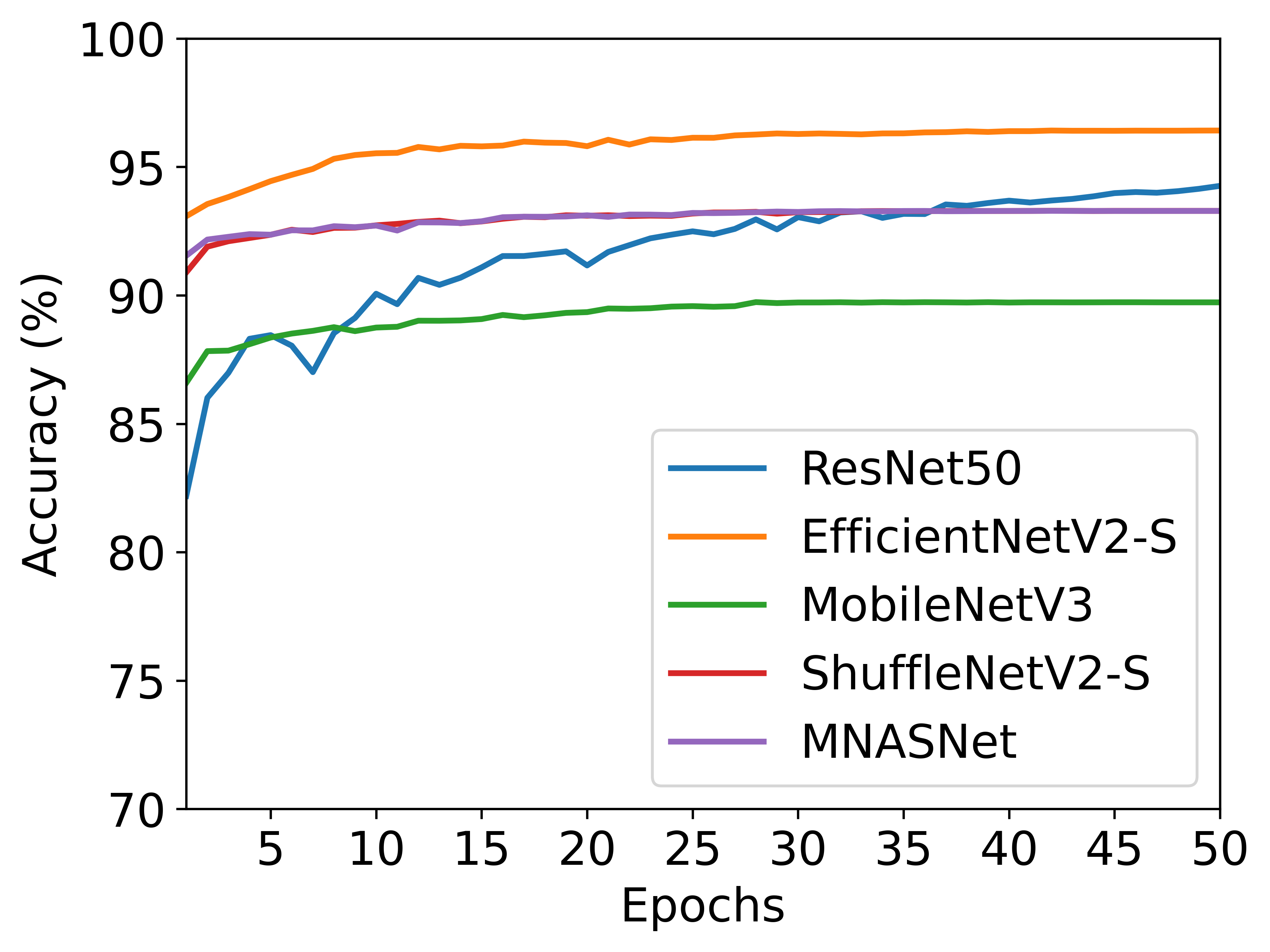}
	}
	\caption{Training accuracy (left) and validation accuracy (right) of ESupNNet over benchmark models.}
	\label{fig:epochs}
\end{figure*}

The average accuracy of ESupNNet for each benchmark model is shown in Fig. \ref{fig:epochs}, for training and validation on the left and right respectively. It is observed that the accuracy stabilises during the first epochs for models using CIFAR-10, while for ImageNet-1K it keeps increasing. The faster convergence of the formers is mostly a product of the batch size used, which is smaller for MobileNetV3, ShuffleNetV2 and MNASNet. An important result is the better performance of ESupNNet for ResNet-50 and EfficientNetV2-S. This can be a result of the amount of information used by the supervising ANN to obtain $R^*$, which might be higher when using 1,000 classes instead of 10. All models are above 90\% in accuracy except for the one supervising MobileNetV3, which is around 89.7\%. Nonetheless, these results backed by cross-validation, offer an insight of the great capabilities of supervising the output logits of a target CNN to detect errors in parameters.

More metrics are further detailed in Table \ref{tab:metrics}. From columns from left to right show accuracy, true positive rate (TPR) or recall, false negative rate (FNR), true negative rate (TNR), false positive rate (FPR), precision and F2-score are reported after the 50 epochs of training. Each cell contains the mean value and the standard deviation calculated from the 10-folds of cross-validation. One of the most important results is the small deviation of all metrics for 10-folds. It proves the statistical significance of the sampling used to obtain each dataset $D$ from the parent dataset $D_p$. For models using CIFAR-10, the recall is higher than precision, which is the opposite for models using ImageNet-1K. In high reliability applications, a false negative is more costly than a false positive, as it is better to act upon an error than to let the error produce a system failure. Thus, for the benchmarks tested, in general ESupNNet provides a reliable way of detecting classification errors for such applications. As an example, when detecting an error it would be advised to reload the parameters and perform an inference again to check discrepancies. In an autonomous navigation application, the cost of this process is very likely to be less than the effect of a misclassification, e.g. crashing an autonomous drone. On the other hand, false positives incur in a cost that might not be negligible in constrained systems. Therefore, the F2-score is also reported to provide a better comparison between recall and false positives, favouring the former while still considering false positives. The F2-score for all benchmarks is above 90\%, which further represents the great error detection of ESupNNet.

The results reported in Table \ref{tab:metrics} show the great capabilities of ESupNNet. Experiments were done in a general manner, with no hyperparameter tuning specific to each target CNN. For specific applications and target CNNs, doing hyperparameter tuning might result in a better performance.

\begin{table*}[]
	\centering
	\caption{Metrics from ESupNNet obtained with cross-validation for K-folds with K=10}
	\label{tab:metrics}
	\begin{tabular}{llllllll}
		& \multicolumn{1}{c}{\textbf{Accuracy (\%)}} & \multicolumn{1}{c}{\textbf{TPR (\%)}} & \multicolumn{1}{c}{\textbf{FNR (\%)}} & \multicolumn{1}{c}{\textbf{TNR (\%)}} & \multicolumn{1}{c}{\textbf{FPR (\%)}} & \multicolumn{1}{c}{\textbf{Precision (\%)}} & \multicolumn{1}{c}{\textbf{F2-score (\%)}} \\ \hline 
		\textbf{ResNet-50}         & 94.26 $\pm$ 0.25      & 92.31 $\pm$ 0.85               & 7.69 $\pm$ 0.85               & 96.21 $\pm$ 0.75               & 3.79 $\pm$ 0.75               & 96.07 $\pm$ 0.71                                       & 93.04 $\pm$ 0.59                     \\ \hline
		\textbf{EfficientNetV2-S} & 96.42 $\pm$ 0.05      & 95.40 $\pm$ 0.18               & 4.60 $\pm$ 0.18               & 97.43 $\pm$ 0.10               & 2.57 $\pm$ 0.10               & 97.38 $\pm$ 0.10                                       & 95.79 $\pm$ 0.13                     \\ \hline
		\textbf{MobileNetV3}      & 89.73 $\pm$ 0.19      & 91.45 $\pm$ 0.18               & 8.55 $\pm$ 0.18               & 88.01 $\pm$ 0.37               & 11.99 $\pm$ 0.37               & 88.41 $\pm$ 0.31                                       & 90.82 $\pm$ 0.14                     \\ \hline
		\textbf{ShuffleNetV2-S}   & 93.29 $\pm$ 0.20      & 93.72 $\pm$ 0.25               & 6.28 $\pm$ 0.25               & 92.85 $\pm$ 0.24               & 7.15 $\pm$ 0.24               & 92.91 $\pm$ 0.22                                       & 93.56 $\pm$ 0.22                     \\ \hline
		\textbf{MNASNet}          & 93.29 $\pm$ 0.13      & 94.18 $\pm$ 0.23               & 5.82 $\pm$ 0.23               & 92.39 $\pm$ 0.17               & 7.61 $\pm$ 0.17               & 92.53 $\pm$ 0.15                                       & 93.84 $\pm$ 0.19                     \\ \hline
	\end{tabular}
\end{table*}

\begin{figure}
	\centering
	\includegraphics[width=\columnwidth]{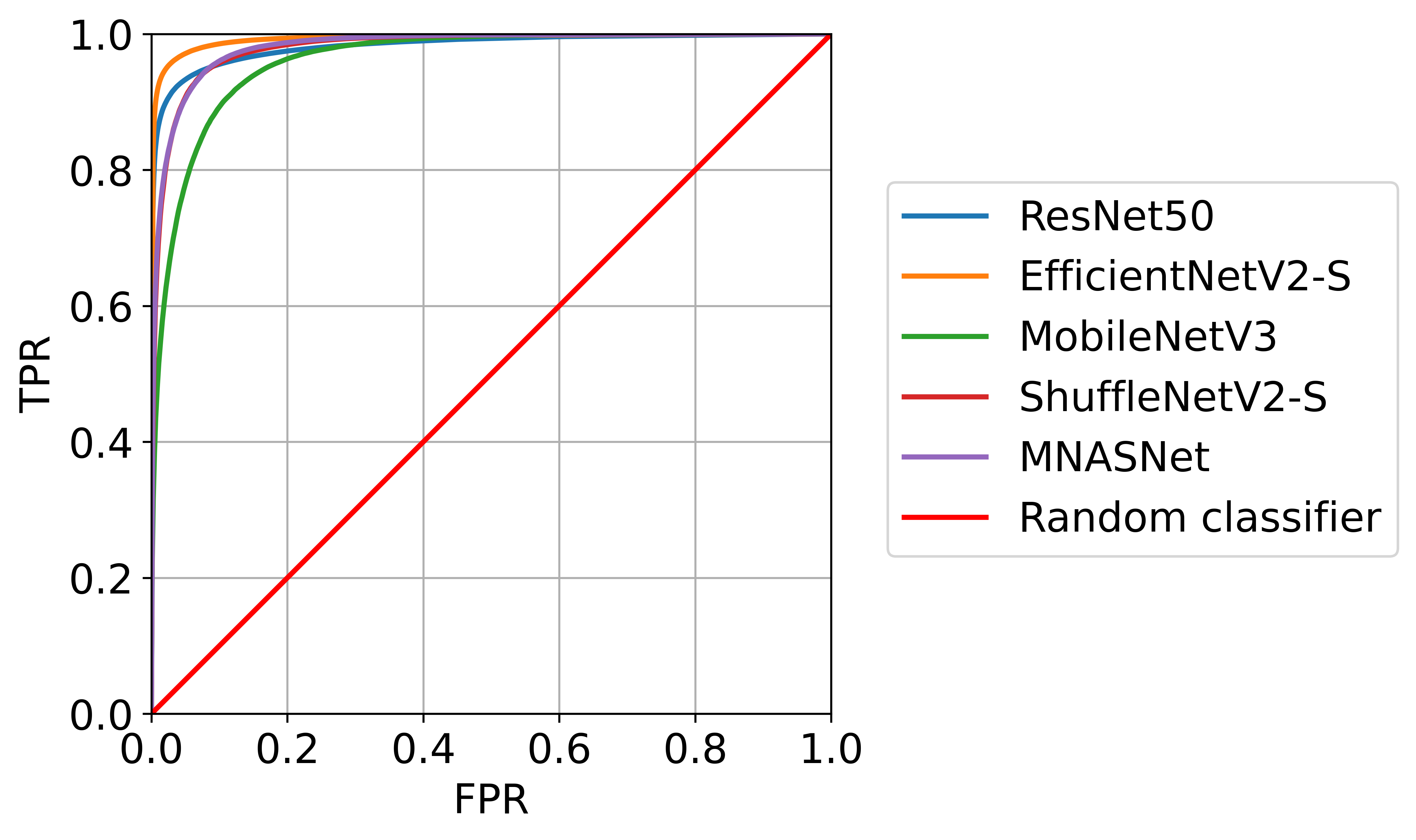}
	\caption{ROC curve of the benchmarks compared to a random classifier}
	\label{fig:auroc}
\end{figure}

Calculating the Area Under the ROC curve (AUROC) provides a balanced way of comparing different solutions. It also allows to select the best threshold for a specific application, as for high reliability applications the cost of false negatives or false positives can be high. We calculated the AUROC to better characterise how the threshold after the sigmoid function of the supervising ANN affects classification. Both the ROC curve and AUROC for all benchmarks are calculated using data from the 10-folds. The average ROC curves are reported in Fig. \ref{fig:auroc} together with that of a random classifier for comparison. A total of 101 threshold points uniformly distributed from 0 to 1 were calculated. For EfficientNetV2, the curve is the one closer to the left, meaning that it has a strong discriminatory capability. It is followed by the curves for ResNet-50, ShuffleNetV2, MNASNet, with the last one being for MobileNetV3. The mean and standard deviation obtained for the AUROC in descending order are the following:

\begin{enumerate}
	\item EfficientNetV2-S: 0.9926 $\pm$ 0.0002
	\item ResNet-50: 0.9825 $\pm$ 0.0011
	\item MNASNet: 0.9816 $\pm$ 0.0007
	\item ShuffleNetV2-S: 0.9806 $\pm$ 0.0007
	\item MobileNetV3: 0.9619 $\pm$ 0.001
	\item Random classifier: 0.5
\end{enumerate}

The first thing to notice are the outstanding values for all benchmarks. The worst result is above 0.96 and the standard deviation is around 0.001. The AUROC is better for models using ImageNet-1K than for those using CIFAR-10. This could mean that with high-dimensional data, the supervising ANN is able to obtain a better class relation $R^*$.

\begin{figure*}[]
	\captionsetup[subfigure]{labelformat=empty}
	\centering
	\subfloat[]{
		\includegraphics[width=0.3\textwidth]{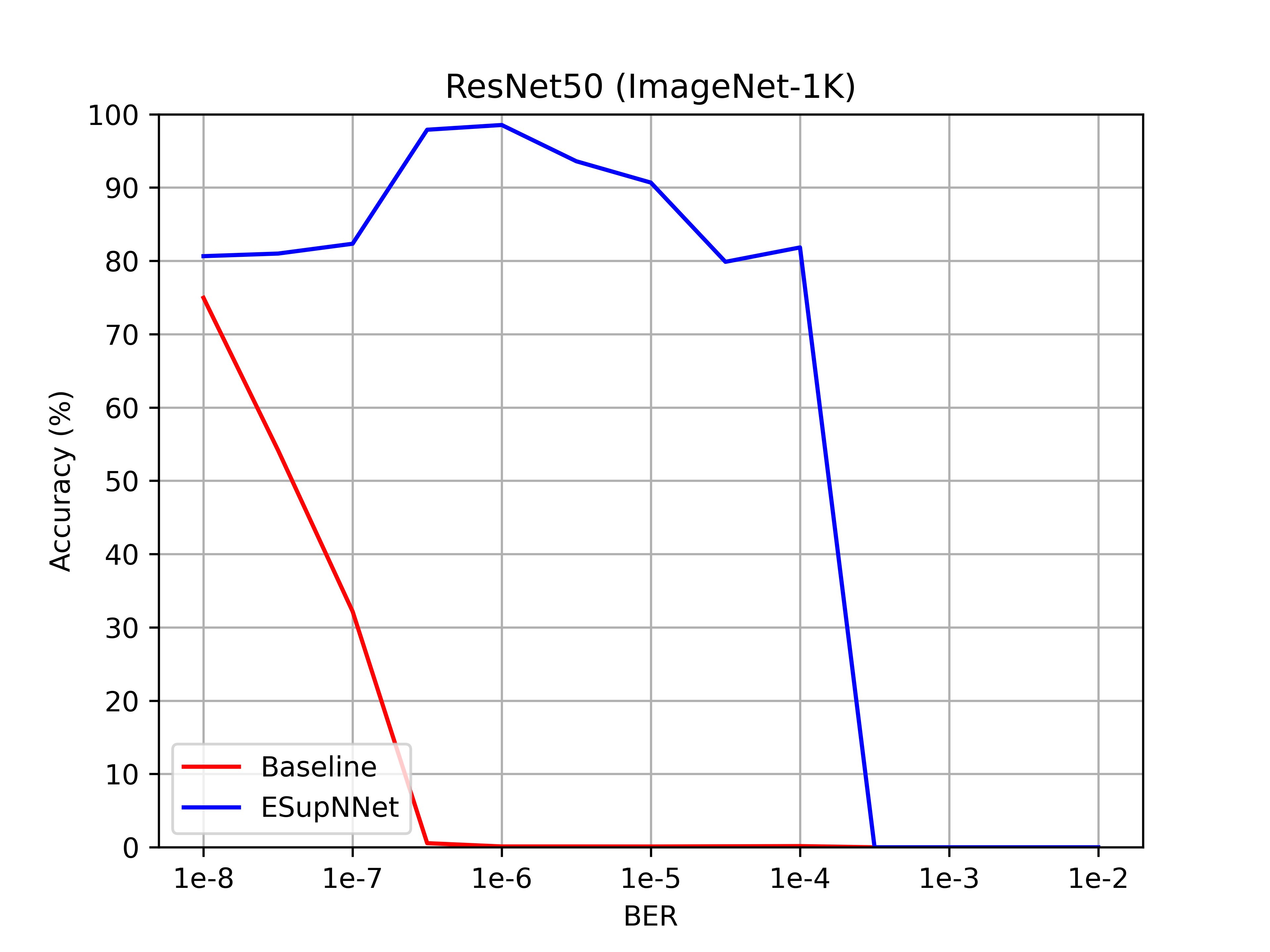}
	}
	\subfloat[]{
		\includegraphics[width=0.3\textwidth]{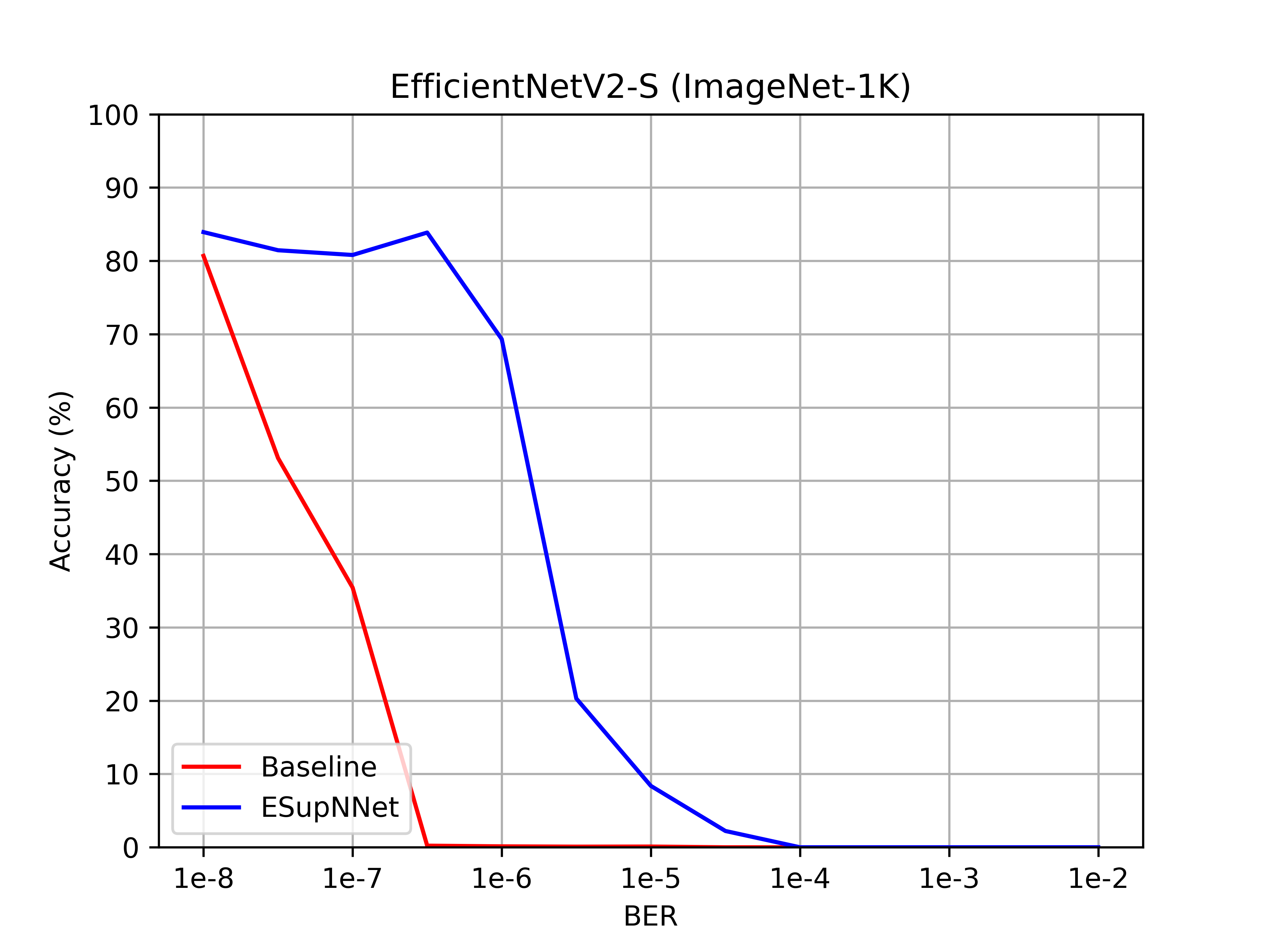}
	}\\
	\subfloat[]{
		\includegraphics[width=0.3\textwidth]{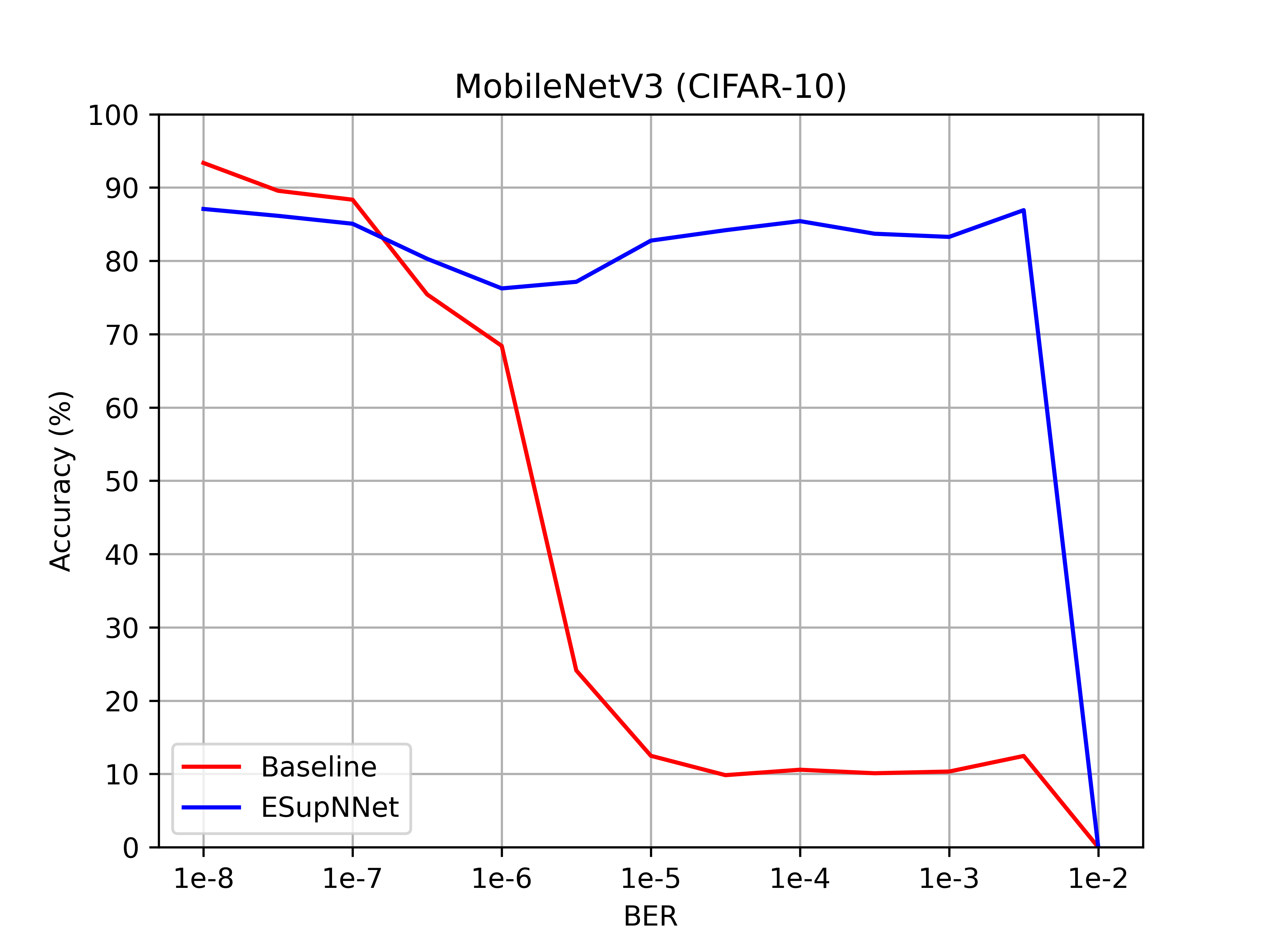}
	}
	\subfloat[]{
		\includegraphics[width=0.3\textwidth]{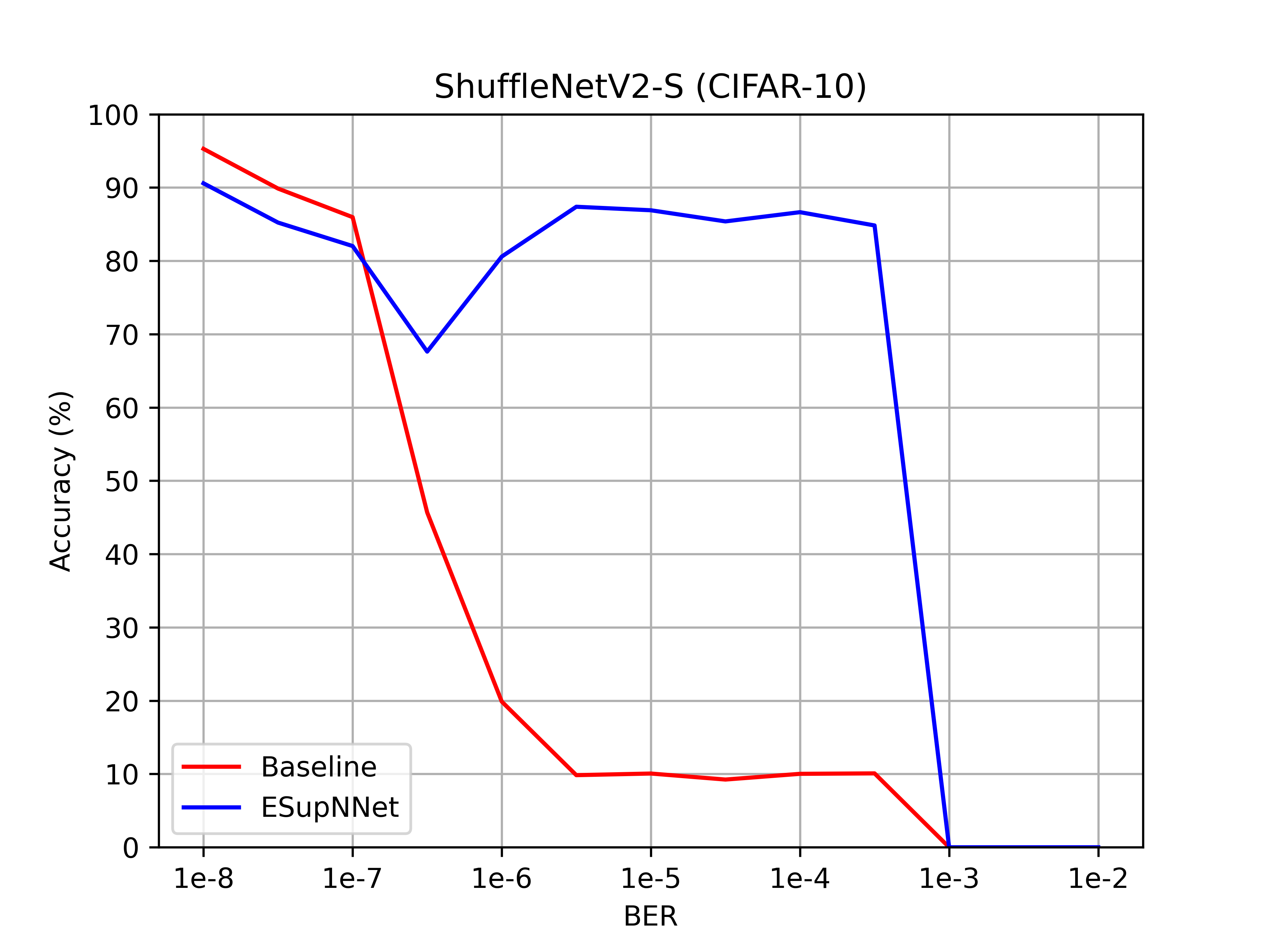}
	}
	\subfloat[]{
		\includegraphics[width=0.3\textwidth]{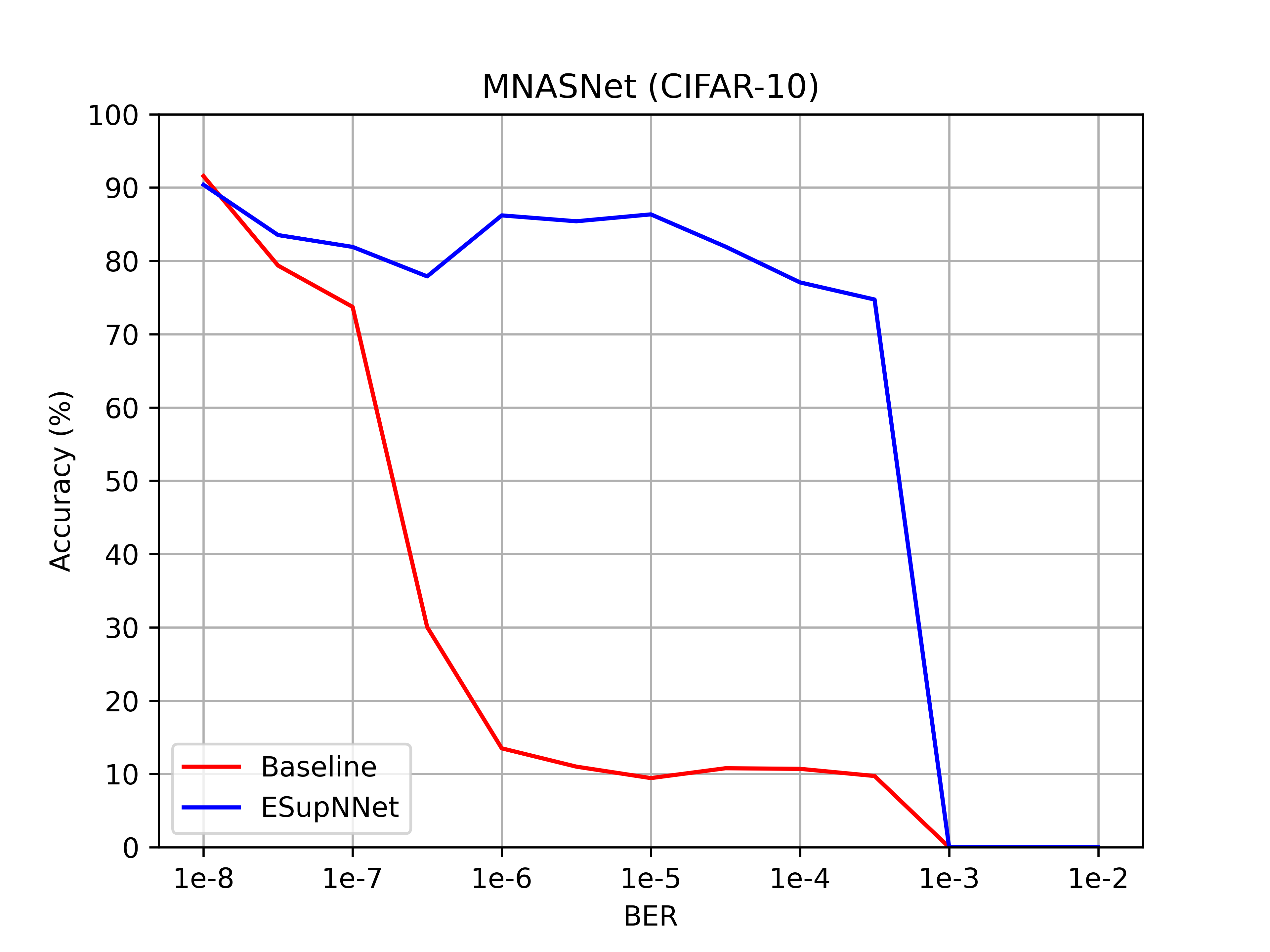}
	}
	\caption{Representation of the fault tolerance of several benchmark models over different Bit Error Rates (BER). Baseline refers to the accuracy of the benchmark CNN without protection. ESupNNet refers to the accuracy of the proposed solution of this work.}
	\label{fig:ber}
\end{figure*}

To analyse the behaviour of ESupNNet over multiple upsets in parameters, we performed simulations for several Bit Error Rates (BER). These simulations were done using the architecture shown in Fig. \ref{fig:arch}, taking the best model of the 10 folds and making inferences of the validation dataset of the target CNN. The BER ranges from $1\times10^{-8}$ to $1\times10^{-2}$, for a total of 13 different BER. For each simulation, a random number of bits are selected and flipped according to the BER and the total number of bits of the parameters. Then, a random set of the validation dataset of the target CNN is processed by the target CNN and the supervising ANN. For each BER, 100 different simulations were done. The results are represented in Fig. \ref{fig:ber}, which shows the Top-1 accuracy of the target CNN (baseline) in red and the proposed architecture (ESupNNet) in blue. NaN or Infinite results are not included, as they can be detected with specific filters and such methods are out of the scope of this work. The sudden descent of accuracy to 0\% for ESupNNet is due to either Baseline or ESupNNet only producing NaN or Infinite results.

The first notable result is that the accuracy of ESupNNet follows a non-monotonic shape, except for EfficientNetV2-S. For ResNet-50 it experiences an increase and decrease around $5\times10^{-7}$ BER. For models using the CIFAR-10 dataset, accuracy drops around $5\times10{-7}$ BER and then increases. This difference between datasets could be produced by the dimensionality of the input data of the supervising ANN. When the dimensionality is high, a better inter class relation $R^*$ can be obtain. As the number of corrupted parameters increases, the inter-class relations of the target CNN are broken and more easily identified by the supervising ANN. We could expect the same behaviour for models using CIFAR-10 dataset, but the dimensionality of the input data of the supervising ANN for these is two orders of magnitude lower. The supervising ANN is not as able to identify when inter-class relations are broken as with high dimensionality models. Nevertheless, the result for EfficientNetV2-S is very different to the one for ResNet-50. The baseline accuracy shows a similar degradation with increasing BER, but the accuracy of ESupNNet rapidly drops in a BER of $1\times10^{-6}$, while for ResNet-50 it happens at a BER of $1\times10^{-4}$.

On the other hand, while the baseline accuracy drops when increasing the BER, the accuracy of ESupNNet does not drop significantly. It only drops to low levels when the target CNN is not able to produce valid results (not NaN or Infinite), which makes it impossible for the supervising ANN to make valid detections at all. These results show that ESupNNet is able to detect classification errors due to MBUs in parameters even for high BER, while maintaining a high accuracy. It shall be noted that the dataset $D$ used to train the supervising ANN does not contain any data regarding MBUs. The supervising ANN, being trained with data of SEUs, is able to detect misclassifications due to MBUs with great accuracy.

\subsection{Comparison with previous works}
In Table \ref{tab:comparison} some data are shown to help compare this work to previous similar works. Data reported is related to resource overhead, if the solution modifies the original model, accuracy without errors of the original model (baseline) and the solution (work), and drop of accuracy depending on BER. For ImageNet-1K dataset, four benchmarks are shown. Stegano-ECC work \cite{stegano_ecc} applies Single Error Correction (SEC) codes to the bits of weight that have an impact on the model inference and uses less relevant bits of the weight to embed such SEC codes. Weight Nulling \cite{weight_nulling} uses a parity bit embedded in weights to detect errors. When an error is detected, the weight is set to a value that minimises distortion. The work in \cite{nonmonotonic} exploits non-monotonic bit sensitivity (NBS in Table \ref{tab:comparison}) in IEEE 754 single-precision of weights to detect or correct SEUs. They propose a correction method that has a 31.25\% resource overhead and a detection one that has zero overhead. For ImageNet-1K dataset, the proposed ESupNNet offers the lower accuracy drop at $1\times10^{-4}$ BER with ResNet-50. For CIFAR-10 dataset, weight nulling exhibits no accuracy drop in BERs inferior to $1\times10^{-4}$. For NBS, although datasets are not the same as the ones used in this work, it is relevant to show the difference in accuracy drop between the correction and detection methods. Furthermore, comparing NBS MobileNetV2 results with this work MobileNetV3, this work has a lower accuracy drop relative to resources overhead. Comparing NBS ResNet-18 with this work ResNet-50, our solution has the lowest accuracy drop.

The works compared in Table \ref{tab:comparison} require a modification of the original model, while ESupNNet does not. For a constrained system, embedding SEC codes in the less important bits of the weights of a model is not feasible. Doing so would mean that the number of bits used to code weights could have been reduced beforehand, so they would be eliminated to reduce hardware. Therefore, these solutions could not be able to be implemented with zero overhead as these bits would be required and use hardware. On the other hand, as ESupNNet does not modify the original model, it can be used the same way for resource constrained and not constrained systems.

\begin{table*}[]
\centering
\caption{Comparison of the proposed ESupNNet with other works.}
\label{tab:comparison}
\begin{tabular}{lccccccccc}
	\hline
	\multicolumn{1}{c}{\multirow{2}{*}{\textbf{WORK}}} & \multirow{2}{*}{\textbf{DATASET}}                        & \multirow{2}{*}{\textbf{MODEL}} & \multirow{2}{*}{\textbf{\begin{tabular}[c]{@{}c@{}}RESOURCES\\ OVERHEAD (\%)\end{tabular}}} & \multirow{2}{*}{\textbf{MODIFICATION}}                     & \multicolumn{2}{c}{\textbf{ACCURACY (\%)}}       & \multicolumn{3}{c}{\textbf{ACC. DROP / BER (\%)}}                       \\
	\multicolumn{1}{c}{}                               &                                                          &                                 &                                                                                        &                                                            & \textbf{Baseline}    & \textbf{Work}        & \textbf{1e-8}        & \textbf{1e-6}        & \textbf{1e-4}        \\ \hline \hline
	\textbf{This work}                                 & \multirow{4}{*}{ImageNet-1K}                             & \multirow{2}{*}{ResNet-50}      & 0.86                                                                                 & NO                                                         & 80.86              & 81.62              & 0.98               & -17.98             & 1.84               \\ \cline{1-1} \cline{4-10} 
	Stegano-ECC \cite{stegano_ecc}                                          &                                                          &                                 & 0                                                                                  & YES                                                     & 75.00              & 75.00              & 0.00               & 0.00               & 45.00              \\ \cline{1-1} \cline{3-10} 
	\textbf{This work}                                 &                                                          & EfficientNetV2                  & 1.03                                                                                 & NO                                                         & 85.24              & 85.52              & 1.60               & 16.20              & 85.52              \\ \cline{1-1} \cline{3-10} 
	Stegano-ECC \cite{stegano_ecc}                                         &                                                          & MobileNetV2                     & 0                                                                                  & YES                                                     & 72.00              & 72.00              & 0.00               & 0.00               & 15.00              \\ \hline \hline
	\textbf{This work}                                 & \multirow{4}{*}{CIFAR-10}                                & MobileNetV3                     & 1.13                                                                                 & NO                                                         & 93.21              & 87.16              & 0.08               & 10.92              & 1.74               \\ \cline{1-1} \cline{3-10} 
	\textbf{This work}                                 &                                                          & ShuffleNetV2                    & 0.32                                                                                 & NO                                                         & 95.12              & 90.40              & -0.18              & 9.82               & 3.76               \\ \cline{1-1} \cline{3-10} 
	\textbf{This work}                                 &                                                          & MNASNet                         & 0.35                                                                                 & NO                                                         & 91.68              & 90.37              & -0.01              & 4.17               & 13.31              \\ \cline{1-1} \cline{3-10} 
	WeightNull. \cite{weight_nulling}                                       &                                                          & Custom                          & 0                                                                         & YES & 89.90              & 89.90              & 0.00               & 0.00               & 0.00               \\ \hline \hline
	NBS \cite{nonmonotonic}                                                & UCMerced                                                 & MobileNetV2                     & \begin{tabular}[c]{@{}c@{}}31.25\\ (0)\end{tabular}                                                                                 & YES                                                     & 97.62              & 97.62              & -                    & -                    & \begin{tabular}[c]{@{}c@{}}0.88\\ (3.44)\end{tabular}                \\ \hline
	NBS \cite{nonmonotonic}                                                & \begin{tabular}[c]{@{}c@{}}NWPU-\\ RESISC45\end{tabular} & ResNet-18                       & \begin{tabular}[c]{@{}c@{}}31.25\\ (0)\end{tabular}                                                                                 & YES                                                     & 94.16              & 94.16              & -                    & -                    & \begin{tabular}[c]{@{}c@{}}3.62\\ (7.79)\end{tabular}                \\ \hline
	& \multicolumn{1}{l}{}                                     & \multicolumn{1}{l}{}            & \multicolumn{1}{l}{}                                                                   & \multicolumn{1}{l}{}                                       & \multicolumn{1}{l}{} & \multicolumn{1}{l}{} & \multicolumn{1}{l}{} & \multicolumn{1}{l}{} & \multicolumn{1}{l}{} \\
	& \multicolumn{1}{l}{}                                     & \multicolumn{1}{l}{}            & \multicolumn{1}{l}{}                                                                   & \multicolumn{1}{l}{}                                       & \multicolumn{1}{l}{} & \multicolumn{1}{l}{} & \multicolumn{1}{l}{} & \multicolumn{1}{l}{} & \multicolumn{1}{l}{}
\end{tabular}
\end{table*}

Results obtained throughout all the experiments and metrics in this work show the great potential of ESupNNet architecture for error detection. It is common to almost all data presented, that results for models using ImageNet-1K dataset are better than for those using CIFAR-10. This can mean that the relation $R^*$ is better approximated to $R$ the higher the dimensionality of the input data. As such, ESupNNet can be used in a general manner to obtain great results, but it might be possible to increase its performance by increasing the dimensionality of the input data of the supervising ANN and doing some hyperparameter tuning. As a consequence of this outcome, it could also be possible to increase its performance by using internal neurons of the target CNN. Also, creating a dataset $D$ that contains information relative to multiple bit upsets could further improve its performance over different BERs. Moreover, the minimal overhead of ESupNNet and its non-intrusiveness over the target CNN makes it an interesting solution to be used with other error protection mechanisms. Lastly, compared to other solutions, our proposed architecture does not modify the target CNN and have a negligible overhead in terms of resources.

\section{Conclusion}
We propose a new form of detecting misclassification errors in CNNs, consisting on observing the inter-class relations of the output logits of the CNN. This relation between classes is induced from the feature extractor and fully connected part of the CNN. No retraining or modification of the original CNN are required and it can even be considered a black-box as long as its output logits are available. We propose an architecture for CNNs called ESupNNet, which is validated with a combination of five different dataset-model benchmarks: ImageNet-1K for ResNet-50 and EfficientNetV2-Small; and CIFAR-10 for MobileNetV3, ShuffleNetV2-Small and MNASNet1\_3. The validation is done rigorously from the creation of the dataset used by ESupNNet to the obtention of its metrics by a K-folds cross-validation with K=10. Many metrics of its performance are provided so it can be fully analysed and compared with other techniques. Experimental results show great performance over all benchmarks, with minimum deviation from each fold. An error-detection accuracy of over 90\%, AUROC above 0.96 and great accuracy for several Bit Error Rates are obtained. It might even be possible to increase its performance with hyperparameter fine tuning. The ease of implementation, low overhead ($\leq$1\%) and performance of ESupNNet, prove that our proposal can compete with state of the art solutions. Moreover, it can potentially be used together with other techniques to create specialised high reliability systems using CNNs.

\bibliographystyle{IEEEtran}
\bibliography{{bibliography/references}}

\vfill

\end{document}